\documentclass[aps,prb,twocolumn,eqsecnum,showpacs]{revtex4}
\usepackage{graphicx}

\usepackage[dvips]{epsfig}

\begin{document}
\title{On the Generation of X-Ray Photon Pairs: A Verification of the Unruh Effect?}
\author{Friedhelm Bell\footnote{Electronic address: friedhelm.bell@bnw4u.de}}
\affiliation{Department f\"ur Physik,
Ludwig-Maximilians-Universit\"at, D-85748 Garching, Germany}

\begin{abstract}
We investigate the production of entangled X-ray photons by
incoherent double Compton scattering (IDCS) which seems to be an
interesting alternative to coherent double Compton scattering used
in earlier work.  Very recently it has been proposed to use the
Unruh effect as a source for photon pairs (R.Sch\"utzhold et
al.,Phys.Rev.Lett.{\bf 100},091301 (2008)). We will discuss this
mechanism in comparison with inverse IDCS.
\end{abstract}

\pacs{03.67.Mn, 42.65.-k, 41.60.-m, 04.62.+v, 42.50.Dv}

\maketitle

\section{Introduction}
The production of correlated two-photon states by parametric down
conversion in nonlinear media is well known  in the visible
optical region \cite{/33/}. Here, down conversion is meant as the
spontaneous decay of a single photon into a pair whose energies
add up to the primary one. The effect is usually described as the
mixing of real photons with vacuum fluctuations resulting in two
real photons. The importance of entangled photons will be realised
if one recognises that they are fundamental incredients of future
quantum computers or quantum crypto\-graphy. Whereas in the
visible optical region rather strong sources for photon pairs are
available the situation for entangled X-ray photons is very
different. Motivated by a proposal of Freund and Levine \cite{/1/}
Eisenberger and McCall \cite{/2/} where the first who realized
X-ray parametric down conversion experimentally. One of the
remarkable features of both the proposal and the experiment was
that the recoil of the photon pair should be overtaken by the
target crystal, thus providing some kind of coherence. Whereas the
Eisenberger-McCall experiment used a conventional X-ray tube as
the primary photon source, the experiment was later repeated with
synchrotron radiation from storage rings \cite{/3/,/4/}, but even
then,  the event rate did not exceed 0.1 photon pairs/s.  Both by
Eisenberger \cite{/2/} and Adams \cite{/4/} the results were
discussed in terms of mixing the incoming X-ray photon with vacuum
fluctuations of the photon field at frequencies of the photon
pair. The high power density of these fluctuations especially at
X-ray energies compensates for the small optical nonlinearity of
the target material, making down conversion observable. Whereas in
these experiments \cite{/2/,/4/} the photon pair was detected by a
coincidence technique we also mention experimental work where only
one of the photons was observed while the other had vanishing
small energy \cite{/5/,/6/}.\\
While the explanation of entangled X-ray photon generation as
described above seems to be influenced by the picture for down
conversion by vacuum fluctuations a more conventional - and for
our case more appropriate - understanding of the effect was
already offered in the proposal of Freund and Levine:``The
incoherent process of double Compton scattering first discussed by
Heitler and Nordheim, in which a photon when interacting with a
relativistic (and hence {\it nonlinear}) electron decays into two
photons of lesser energy, is well known. We discuss here an
analogous coherent phenomenon, the spontaneous parametric decay of
X-rays. This process is related to double Compton scattering in
the same way that ordinary Bragg diffraction relates to ordinary
Compton scattering.'' \cite{/1/}.\\
In the following we discuss incoherent double Compton scattering
(IDCS) as a source for X-ray photon pairs. It is hoped that the
effieciency of this process is  larger than for coherent double
Compton scattering (CDCS), since stringent conditions for defining
photon momentum vectors necessary to fulfill the phase matching
condition in CDCS do not apply in IDCS. We shall also connect our
analysis with recent publications on a very different source for
correlated X-ray photons: pair production by the Unruh-Hawking
effect in a non-inertial reference frame \cite{/7/,/7b/,/8/}.
While this effect by itself is very important in quantum gravity
it might also have relevance to our theme.  Finally, we mention
that IDCS plays an important role in astrophysics since it was
realised  that it
can become the main source for soft photons in astrophysical
plasmas with low
baryon density \cite{/9/}.\\
Last not least we indicate useful applications of entangled X-ray
photons as in two photon interferometry. Instead of making a
single photon probe two interfering paths, one might also employ
two-photon states in interferometry. We stress that in this case
an interference pattern will be observed even if both photons have
different energies and arrive at different detectors (fourth-order
interference experiments \cite{/31/}), thus allowing phase
shift observations at different energies $\omega_1$ and
$\omega_2$. This, of course,  implies mutual coherence of both
photons. Since quantum effiency of X-ray detectors is
close to 100\% the visibility of higher order interferences to
test spin-free Bell`s inequalities of quantum theory is at maximum
\cite{/31/}. The same holds for Einstein-Podolsky-Rosen
experiments where any defiency of quantum effiency is equivalent
to influences of external degrees of freedom on pure quantum
states.

\section{Incoherent double Compton scattering}
After an order of magnitude calculation by Heitler and Nordheim
\cite{/11/} Mandl and Skyrme \cite{/12/} were the first who
calculated within the framework of quantum electrodynamics the
differential cross section for IDCS exactly. These results have
been extended and intensively discussed in the book of Jauch and
Rohrlich \cite{/13/}.
\begin{figure}[!ht]
\includegraphics[width=4.8cm,clip]{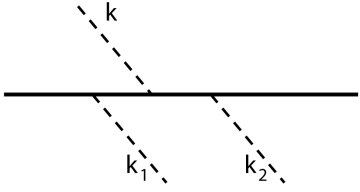}
\label{Fig1} \caption{Feynman diagram for double Compton
scattering}
\end{figure}
Fig.~1 shows one of the 3! equivalent third order Feynman diagrams
for IDCS\cite{/13/}. We note that these diagrams are equivalent
to the six components of the second order susceptibility
calculated in nonlinear optics \cite{/34/}. In the following we
use socalled natural units, i.e. $\hbar = m = c = 1 $. As shown in
Fig.~1 IDCS is a process where an incoming photon with 4-vector $k
= ( \omega, {\it {\bf k}})$ is scatterd at an electron with $p =
$($ \gamma , \gamma $\mbox{\boldmath $\beta$}) resulting in two
final photons with $k_1 = ( \omega_1 , {\it {\bf k}}_1 )$ and $k_2
= ( \omega_2 , {\it {\bf k}}_2)$ and the final electron with $p' =
(\gamma' , \gamma' $\mbox{\boldmath $\beta'$}). Kinematics demand
\begin{eqnarray}
p' + k_1 + k_2 - p - k = 0
\label{eq(1)}
\end{eqnarray}
We use the signature $(-+++)$, i.e. $p^2 + 1 = 0$. The calculation
holds for the lab frame and  $\gamma = (1 - \beta^2)^{-1/2}$.
Replacing the initial electron 4-vector by $p = (M,{\it {\bf 0}})$
and the final one by $p' = ( \sqrt{M^2 + g^2} ,  {\it {\bf g}} )$
, where M is the target mass and ${\bf g}$  a reciprocal lattice
vector, one obtains the kinematics for CDCS.\\
The differential cross  section is given for the emission of
photons 1 and 2 into solid angles $d \Omega_1 ' $  and $d \Omega_2
' $ in the direction $(\theta_1 ' , \varphi_1 ' )$  and $(\theta_2
' , \varphi_2 ' )$. These angles refer to a polar coordinate
system whose axis is the incident electron direction. The azimuths
$\varphi'$ are conveniently measured from the plane of incidence
formed by the vectors $ {\bf k}$  and \mbox{\boldmath $ \beta$} .
We keep the notation of ref.~15. Now, let $\theta_{1}$ and
$\theta_2$  be the angles between the directions of the outgoing
photons and the incoming one and $\theta_{12}$  that between the
pair, then
\begin{eqnarray}
\cos\theta_1 &=& \cos\alpha \cos\theta_1 ' + \sin\alpha \sin\theta_1 ' \cos\varphi_1 ' \\
\cos\theta_2 &=& \cos\alpha \cos\theta_2 ' + \sin\alpha \sin\theta_2 ' \cos\varphi_2 ' \\
\cos\theta_{12} &=& \\ \nonumber & & \cos\theta_1 ' \cos\theta_2 '
+ \sin\theta_1 ' \sin\theta_2 ' \cos( \varphi_1 ' - \varphi_2 ' )
\label{eq(2)}
\end{eqnarray}
holds, where $\alpha$  is the angle between $ {\bf k}$  and
\mbox{\boldmath $ \beta$}. For a given energy $\omega_1$ of photon
1 the energy  $\omega_2 $ of photon 2 is obtained from
(\ref{eq(1)})
\begin{eqnarray}\label{eq(3)}
\lefteqn{\omega_2 = } \\ \nonumber & & \frac{ \gamma \omega ( 1 -
\beta \cos\alpha ) - \gamma \omega_1 (1- \beta \cos\theta_1 ' ) -
\omega_1 \omega (1 - \cos\theta_1 )}{\gamma (1 - \beta
\cos\theta_2 ' ) + \omega (1- \cos\theta_2) - \omega_1 (1 -
\cos\theta_{12})}
\end{eqnarray}
Then, the triple differential cross section for IDCS reads
\begin{widetext}
\begin{eqnarray}
\frac{d^3 \sigma_{DC} }{d \omega_1 d \Omega_1 ' d \Omega_2 ' }= 
\frac{\alpha_{QED}}{(4 \pi)^2 } \ r_0^2 \ \frac{X \omega_1
\omega_2}{\gamma \omega ( 1 - \beta \cos\alpha )[\gamma (1- \beta
\cos\theta_2 ') + \omega (1 - \cos\theta_2 )  - \omega_1 (1 -
\cos\theta_{12} )]} \label{eq(4)}
\end{eqnarray}
\end{widetext}
with the cross section function
\begin{eqnarray}
\lefteqn{X = 2 (ab - c ) [ (a+b) (x+2) -(ab-c)-8]- }\\ \nonumber &
& 2x(a^2 + b^2 ) -8c + \frac{4}{AB} [(A+B) (x^2 + x )- \\
\nonumber & & (aA + bB)(2x + z (1-x)) + \\ \nonumber & & x^3
(1-z)+2zx] -2\rho [ab + c (1-x)] \label{eq(5)}
\end{eqnarray}
The abbreviations read
\begin{eqnarray}
\lefteqn{a = \sum\limits_{i=1}^{3} \ \frac{1}{\kappa_i} \ ; b =
\sum\limits_{i=1}^{3} \ \frac{1}{\kappa_i '} \ ; c =
\sum\limits_{i=1}^{3} \ \frac{1}{\kappa_i \kappa_i '}} \\
\nonumber
\lefteqn{x = \sum\limits_{i=1}^{3} \ \kappa_i \ ; z =  \sum\limits_{i=1}^{3} \ \kappa_i \kappa_i '} \\
A &=& \prod^3_{i=1} \kappa_i ; \ B =  \prod^3_{i=1} \kappa'_i ; \
\rho =  \sum\limits_{i=1}^{3} \left( \frac{\kappa_i }{\kappa_i '} \ + \ \frac{\kappa_i '}{\kappa_i} \right)
\label{eq(6)}
\end{eqnarray}
with
\begin{eqnarray}
\kappa_1 = -p \cdot k_1 =  \gamma \omega_1 (1 - \beta \cos\theta_1 ')\\
\kappa_2 =  -p \cdot k_2 =  \gamma \omega_2 (1 - \beta \cos\theta_2 ')\\
\kappa_3 = + p \cdot k = - \gamma \omega (1 - \beta \cos\alpha)
\label{eq(7)}
\end{eqnarray}
and
\begin{eqnarray}
\lefteqn{\kappa_1 '= +p' \cdot k_1 = }\\ \nonumber & & -\kappa _1 - \omega \omega_1 (1-\cos\theta_1) + \omega_1 \omega_2  (1-\cos\theta_{12})\\
\lefteqn{\kappa_2 '= +p' \cdot k_2 = }\\ \nonumber & & - \kappa_2 - \omega \omega_2 (1-\cos\theta_2) + \omega_1 \omega_2  (1-\cos\theta_{12})\\
\lefteqn{\kappa_3 '= -p' \cdot k = }\\ \nonumber & & - \kappa_3 -
\omega \omega_1 (1-\cos\theta_1) - \omega \omega_2
(1-\cos\theta_{2}) \label{eq(8)}
\end{eqnarray}
It is $r_0 = 2.82  \ \cdot 10^{-15}$m   the classical electron
radius and $\alpha_{QED} = 1/137$  the fine structure constant.
The triple differential cross section of IDCS (eq.~(\ref{eq(4)}))
has been verified quantitatively by several authors \cite{/14/}.
In the following we will apply this cross section to special
cases.

\section{Fixed targets}
In case of a solid state target we assume that the electrons are
at rest, i.e. $\beta = 0$. This implies that the angles $(\theta_1
' , \varphi_1 ' )$   and $(\theta_2 ' , \varphi_2 ' )$   become
unimportant and the cross section will be differential with
respect to $(\theta_1  , \varphi_1 ) $   and  $(\theta_2  ,
\varphi_2 )$.  The angle between the two photons becomes
\begin{eqnarray}
\cos\theta_{12} = \cos\theta_1  \cos\theta_2 + \sin\theta_1  \sin\theta_2  \cos(\varphi_1 - \varphi_2)
\label{eq(9)}
\end{eqnarray}
We first calculate the cross section in case of hard X-rays, i.e.
$\omega$~=~100~keV. At storage rings of the 3$^{rd}$ generation
photon fluxes of about $10^{12}$  photons/s can be achieved with a
relative monochromaticity of less than 1\% \cite{/16/}.
\begin{figure}[tb]
\includegraphics[width=0.45\textwidth,clip]{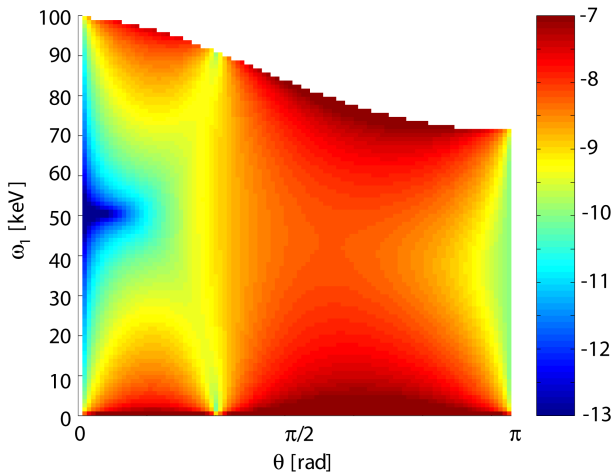}
\label{Fig2} \caption{(Color) Isodensity plot of the triple
differential cross section for double Compton scattering (fixed
target) as a function of $\omega_1 $ and the scattering angle
$\theta$. The scale is logarithmic, i.e.~the number -10 at the
colour scale means $10^{-10}$b/keV$\cdot$sr$^2$. The plot holds
for $\theta_1 = \theta_2 = \theta$, $\varphi_1 = 0$ and
$\varphi_2 = \pi$. The emission of both photons is symmetric with
respect to the direction of the incomming photon,whose energy is
$\omega$= 100 keV}.
\end{figure}
Fig.~2 shows the triple differential cross section for $\varphi_1
= 0$ , $\varphi_2 = \pi$ and $\theta_1 = \theta_2 = \theta$  as a
function of  $\theta $. As a typical value for the cross section
we get for $\omega_1 = \omega_2 $= 42 keV  and $\theta$ = 2 rad
$d^3 \sigma_{DC}$ = 8~nb/keV$\cdot$sr$^2$. It is readily seen that
for either $\omega_1 = 0$  or $\omega_2 = 0$ -the upper limit for
$\omega_1$  -the cross section diverges reflecting the well known
infrared catastrophe if QED cross sections are calculated
perturbatively from S-matrix theory. It was Feynman who first
showed that when radiative corrections to single Compton
scattering and the infrared divergence of double Compton
scattering are calculated on the same footing both divergences
cancel \cite{/17/}. It is also seen that the cross section
vanishes when both photons are emitted exactly in forward
direction. It is easily demonstrated that in this case the
kinematics of (\ref{eq(1)}) cannot be fulfilled. We note that the
maximum energy $\omega_{1 max}$  for photon 1, which corresponds
to $\omega_2 = 0$ , is identical with the energy $\omega'$  for
single Compton scattering
\begin{eqnarray}
\omega_{1max} = \omega' = \frac{\omega}{1 + \omega (1 - \cos\theta_1)}
\label{eq(10)}
\end{eqnarray}
independently of the setting of photon detector 2. An equivalent
conclusion holds for $\omega_{2max}$. Only in case that both
photons are emitted in the {\it same} direction ($\theta_{12} =
0$) it is seen from (\ref{eq(3)}) that the sum $\omega_1 +
\omega_2 $ of  both photon energies is identical to $\omega'$  at
that angle.\\
Setting photon detectors at the most intense cross sections,
i.e.~at $\theta$ = 2~rad with solid angles $\Delta \Omega = 3
\cdot 10^{-2}\,$sr, assuming a 100~$\mu$m thin Al target and a
detected energy bandwidth of 5\% we end up with photon-pair
production rates shown in Fig.~3.
\begin{figure}[tb]
\includegraphics[width=0.4\textwidth,clip]{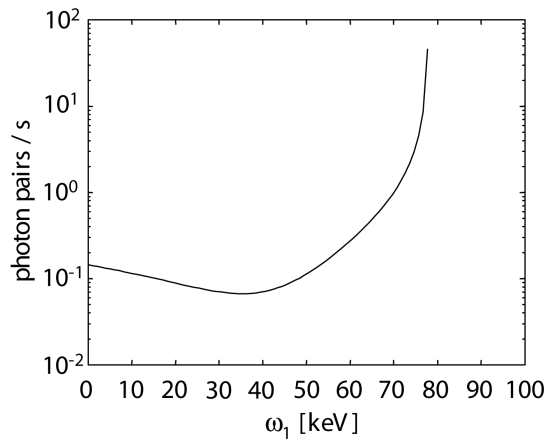}
\label{Fig3} \caption{The pair production rate for the cross
section of Fig.~2 as a function of photon energy.}
\end{figure}
Between 20 and 70 keV the production rate increases from 0.1 to 1
pairs/s, which is comparable to or a factor 10 larger than that
from CDCS mentioned in the Introduction. Even stronger two-photon
sources might be available at forthcoming X-ray Free Electron
Lasers (XFEL). The European XFEL project envisages for the SASE 1
device an average photon flux of $4 \ \cdot \ 10^{16}$  photons/s
at an energy of $\omega = 12.4$~keV \cite{/18/}.
\begin{figure}[tb]
\includegraphics[width=0.4\textwidth,clip]{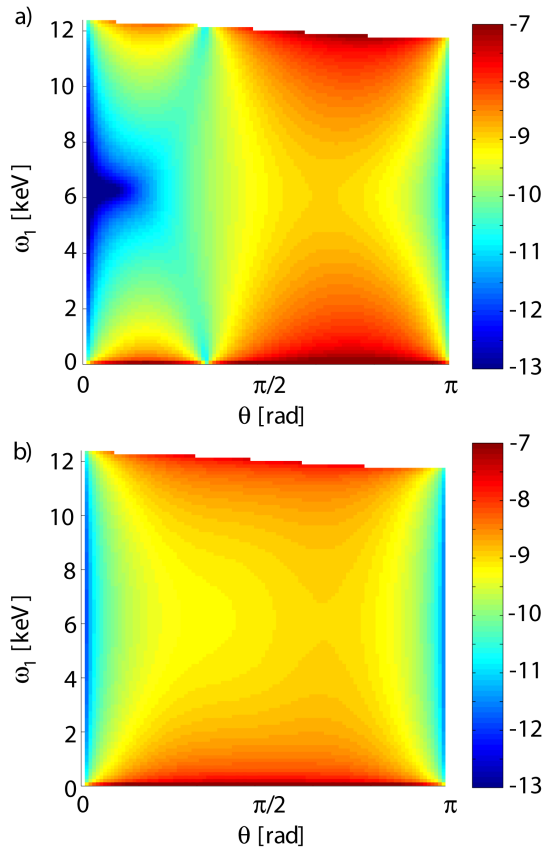}
\label{Fig4} \caption{(Color) (a) The same as Fig.~2 except for
$\omega$=12.4 keV. (b) The same as in (a), but now for $\varphi_1
= \varphi_2 = 0$, i.e. both photons are emitted in the same
direction.}
\end{figure}
Again, Fig.~4a shows the cross section for $\theta_1 = \theta_2 =
\theta $  and  $\varphi_1 = 0 \ , \varphi_2 = \pi$ (which, in the
language of nonlinear optics, is the one-mode case) and $\varphi_1
= \varphi_2 = 0$ (Fig.~4b., i.e. the two-mode case) respectively.
In the latter case, both photons are emitted in the same
direction. The cross sections are roughly an order of magnitude
smaller than those of Fig.~2.\\
As in Fig.~3 we have calculated the pair production rate but now
for a 10~$\mu$m thin Al target to avoid strong photo-absorption.
For the cross section of Fig.~4b and $\theta$= 2.1~rad the pair
production rate is plotted in Fig.~5.
\begin{figure}[tb]
\includegraphics[width=0.4\textwidth,clip]{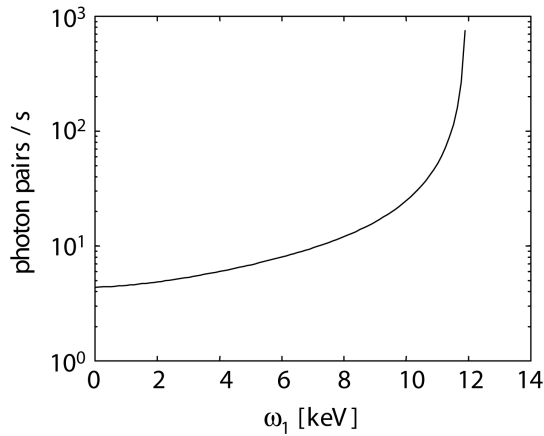}
\label{Fig5} \caption{The pair production rate for the cross
section of Fig.~4b as a function of photon energy.}
\end{figure}
Between 4 and 10 keV it changes from 6 to 30 pairs/s. A remark
seems to be indicated about the cross section when both photons
are emitted in the same direction. The triple differential cross
section describes the probability that the two photons are emitted
into arbitrary close but nevertheless different solid angles which
should not overlap. But if it is stressed that both photons are
emitted exactly into the same solid angle the process should be
describable by a double differential cross section. Going back to
the definition of the cross sections we argue that in this special
case $d^2 \sigma_{DC} / d \omega_1 d \Omega = 4 \pi d^3
\sigma_{DC} / d \omega_1 d \Omega_1 d \Omega_2 $ For the production
rate of Fig.~5 this would mean an increase by a factor $4 \pi / 3
\cdot 10^{-2} = 400$.\\

\section{Inverse Double Compton Scattering}
In this chapter we consider the scattering of low energy optical
laser photons at high energetic electrons with $\gamma >> 1$ where
two photons are generated in the exit channel. The reason is
threefold: i) by this inverse IDCS extremely hard photon pairs
even in the MeV range can be produced, ii) a comparison with
Unruh-Hawking radiation mentioned in the Introduction can be made,
and iii) a connection to undulator radiation will be drawn.\\
i) In case of inverse IDCS one derives from eq.(~\ref{eq(3)}) the
maximum photon energy $\omega_{1max}$  (i.e. $\omega_2 = 0$) in
detector~1
\begin{eqnarray}
\omega_{1max} = \frac{\gamma \omega_L (1 - \beta \cos\alpha)}{\gamma (1 - \beta \cos\theta_1 ' ) + \omega_L (1 - \cos\theta_1)}
\label{eq(13)}
\end{eqnarray}
where $\omega_L $ is the laser photon energy. For $\gamma >> 1$
and with the approximation $\cos\theta_1 \simeq \cos\alpha$,
(\ref{eq(13)}) can be written
\begin{eqnarray}
\omega_{1max} = \frac{\omega_m}{1 + (\theta_1 ' / \theta_0 ' )^2 }
\label{eq(14)}
\end{eqnarray}
where $\omega_{m} = \gamma x / (1+x)$   and  $\theta_0 ' = \sqrt{1
+ x } / \gamma$. It holds $x = 4 \gamma \omega_L \cos^2 (\alpha_0
/ 2)$ with  $\alpha_0 = \pi - \alpha$. Eq.(~\ref{eq(14)}) is
identical with the scattered photon energy of inverse single
Compton scattering \cite{/19/}. To calculate the pair production
rate as in chapter III one has to evaluate the luminosity $L_{e
\gamma}$ for realistic conditions. Assuming for both the electron
and photon bunch perfect overlapping in space and time, the
luminosity per bunch crossing and a head-on collision reads
\begin{eqnarray}
L_{e \gamma} = \frac{N_e N_{\gamma}}{2 \pi ( \sigma_{te} ^2 + \sigma_{t \gamma} ^2)}
\label{eq(15)}
\end{eqnarray}
$N_e$ and $N_{\gamma}$ are the total electron and photon numbers
within the bunches, and $\sigma_{te} , \sigma_{t \gamma}$ the rms
values of the transverse extensions of the bunches. For realistic
numbers we have adopted a scenario which has been proposed for
future table-top FEL`s \cite{/21/}. Then, the number of
electrons/bunch is about 1 nC or $N_e = 10^{10}$. Assuming a laser
with $\omega_L = 2.5$~eV, an intensity $I_L = 10^{18}$W/cm$^2 $, a
pulse duration $\tau_L $ = 50~fs and a matching of the transverse
extensions of both bunches one arrives at a  luminosity/ electron
\begin{eqnarray}
L_{e \gamma} / N_e = \frac{ I_L \tau_L}{4 \hbar \omega_L} \ = 0.06
\ b^{-1} \label{eq(16)}
\end{eqnarray}
\begin{figure}[!ht]
\includegraphics[width=0.4\textwidth,clip]{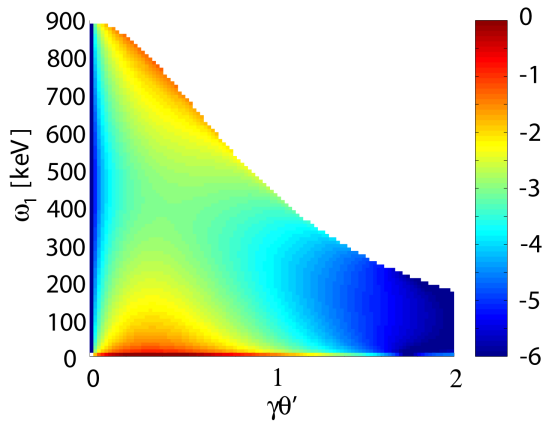}
\label{Fig6} \caption{(Color) Isodensity plot of the photon-pair
yield Y (see text) by inverse double Compton scattering in case of
$\omega_L$ = 2.5 eV and $\gamma$=300. It holds  $\theta_1 ' =
\theta_2 ' = \theta'$ and $\varphi_1 ' = 0, \varphi_2 ' = \pi$.
The scale is logarithmic, i.e. the number -3 at the color scale
means $10^{-3}$ pairs/keV$\cdot$sr$^2$.}
\end{figure}
In Fig.~6 we have plotted the triple differential yield/electron
$Y = d^3 \sigma_{DC} L_{e \gamma} / N_e$~[pairs/keV$\cdot$sr$^2$]
for $\gamma = 300$ and $\theta' _1 = \theta' _2 = \theta' $ ,
$\varphi' _1 = 0, \varphi' _2 = \pi$. The maximum energy in
forward direction amounts to $4 \gamma^2 \omega_L = $ 900~keV.
Employing the most intense yield Y - corresponding to a cross
section of 8~mb/keV$\cdot$sr$^2$ - at an angle $\gamma \theta' =
0.6$, an energy resolution of 5\% for the symmetric case $\omega_1
= \omega_2 = \omega_{1max} /2 = $160~keV and solid angles $\Delta
\Omega' _1 = \Delta \Omega' _2 = 10^{-6}sr$, we obtain a pair
production rate
\begin{eqnarray}
Y N_e \Delta \Omega_1 ' \Delta \Omega_2 '  \Delta \omega_1 = 4 \cdot 10^{-5} \ \rm{pairs} / \rm{pulse}
\label{eq(17)}
\end{eqnarray}
Going to the asymmetric case, i.e. $\omega_1 \neq \omega_2$, one
can easily gain an order of magnitude in yield. We also mention
that for a head-on collision the two-photon pulse duration is
$\sigma_{le} / c $ with $\sigma_{le}$  as the rms length of the
electron bunch, and is thus independent of the laser pulse
duration $\tau_L$.\\
In order to get a feeling about the magnitude of differential
cross sections we will compare single Compton cross sections with
those for double Compton scattering. Due to the finite pulse
duration the laser has a rather strong energy broadening $\Delta
\omega_L = 1 /   \tau_L$. Thus one finds for the double
differential cross section for single Compton scattering
\begin{eqnarray}
\frac{d^2 \sigma_S }{d \omega d \Omega}= \frac{d \sigma_S
}{d\Omega} G (\omega)
\label{eq(18)}
\end{eqnarray}
where $d \sigma_S / d \Omega$  is the cross section for inverse
single Compton scattering \cite{/13/} (unpolarized incomming
photons):
\begin{eqnarray}
\lefteqn{\frac{d \sigma_S }{d \Omega} = \frac{r_0 ^2 }{2} \left( \frac{\omega' }{\kappa} \right)^2}\\
\nonumber & & \times \left[ \frac{\kappa'}{\kappa} +
\frac{\kappa}{\kappa'} + 2 \left( \frac{1}{\kappa'} -
\frac{1}{\kappa} \right) + \left( \frac{1}{\kappa} -
\frac{1}{\kappa'} \right)^2 \right]
\end{eqnarray}
with
\begin{eqnarray}
\omega' = \frac{\gamma \omega_L (1 - \beta \cos \alpha)}{\gamma (1 - \beta \cos \theta' ) + \omega_L (1 - \cos \alpha)}
\end{eqnarray}
and  $\kappa = p \cdot k = - \gamma \omega_L (1 - \beta \cos
\alpha )$, $\kappa' = p \cdot k' = - \gamma \omega' (1 - \beta
\cos \theta')$. In case of a Gaussian bunch the spectral function
$G ( \omega)$ reads
\begin{eqnarray}
G (\omega) = \frac{\omega_L \tau_L }{ \omega' \sqrt{2 \pi } } \
\rm{exp} \ \left( - \frac{1}{2}  (\omega_L \tau_L )^2 (\omega /
\omega' -1 )^2 \right)
\end{eqnarray}
On the other hand one obtains the double differential cross
section for double Compton scattering in case where one is not
interested in the second photon
\begin{eqnarray}
\frac{d^2 \sigma_D}{d \omega_1 d \Omega_1 } = \int\limits^{2 \pi}_0 d \varphi' _2 \int\limits^{+1}_{-1} d \cos \theta'_2 \ \frac{d^3 \sigma_D }{d \omega_1 d \Omega_1 d \Omega_2}
\end{eqnarray}
For the laser and electron bunch data given above the double
differential cross sections in case of head-on collision have been
plotted in Fig.~7. The left side of the figure holds for single,
the right one for double Compton scattering. We mention that the
single Compton peak might be further broadened
due to the finite emittance of the electron bunch.\\

ii) A very interesting method for the production of entangled
X-ray photons has been proposed by Sch\"utzhold et al.
\cite{/7/,/7b/}. Accelerated electrons can convert virtual quantum
vacuum fluctuations into real particle {\it pairs}  via
non-inertial scattering which can be understood as a signature of
the Unruh effect \cite{/23/}. In a sense, Unruh radiation might be
called a sister to Hawking radiation \cite{/24/,/25/}. In 1974
Hawking\cite{/24/} had shown that due to quantum fluctuations
black holes, while embedded in a thermal bath with temperature
$T_{UH}$, may evaporate particle pairs. Unfortunately, a
quantitative comparison between the two-photon yields of the
Unruh-Hawking effect and inverse IDCS is difficult to make since
at least for the differential yields no exact numbers  are given in Fig.1 of
ref.9. But what ever the outcome of such a comparison
would be one seems to await a problem: either the Unruh yield is
considerably smaller than that from IDCS, then it becomes
difficult to detect this effect since it belongs kinematically to
the same region as IDCS. 
\begin{figure}[!ht]
\includegraphics[width=0.4\textwidth,clip]{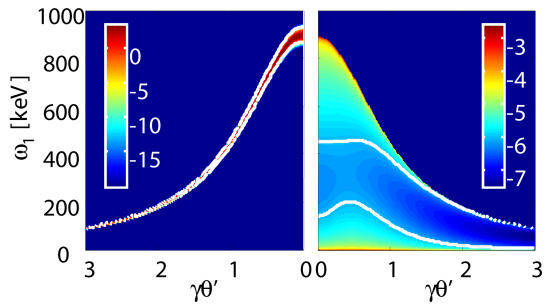}
\label{Fig7} \caption{(Color) Isodensity plot of the double
differential cross section for single (left side) and double
(right side) Compton scattering. Again, the scale is logarithmic,
i.e., the number -6 at the color code means
$10^{-6}$b/keV$\cdot$sr. Note that the color codes for both sides
are rather different. The white contour lines in both sides
correspond to $10^{-6}$b/keV$\cdot$sr. If the cross sections are
multiplied by the luminosity of eq.(~\ref{eq(16)}) one obtains the
yield pairs/(keV$\cdot$sr$\cdot$electron). The figure is drawn in
such a way that a qualitative comparison with Fig.1 of
Sch\"utzhold et al.\cite{/7b/} can be made.}
\end{figure}
This is also demonstrated by a comparison
of Fig.~7 with Fig.1 of Sch\"utzhold et al.~\cite{/7b/}. If the
yield is stronger one has to explain the quantitative agreement
(within a few percent) of experimental data \cite{/14/} with the
theory for IDCS\cite{/12/}. Or, finally, given about the same
yield for both effects, the interesting question arises wether the
physical backgrounds of Unruh-Hawking radiation and IDCS are
the same. If this proves right-and, in fact, such a possibility
has been indicated in ref.~15 of Sch\"utzhold et
al.\cite{/7b/}-the remarkable situation exists, that a signature
for what later was called the Unruh effect had been verified
experimentally 20 years before Unruh
published his seminal paper in 1976 \cite{/23/}. It might also be enlightening to cite a judgement by A. Ringwald \cite{Ring}: ``Whether ultimately one will call this a verification of the Unruh effect or just basic quantum field theory (QED) is a matter of taste or linguistics.'' Adopting this view we recognize a minor similarity only between Fig.1 of ref.9 and our Fig.7. But we also note severe doubts of several authors, whether Unruh radiation exists at all\cite{Ford}. \\ 
iii) Finally, we mention that it is well known that there exists a
strong relationship between an electromagnetic undulator (EMU)
like a laser wave and a magnetic undulator (MU) with magnetic
field strength  $B_U$ used as an insertion device at synchrotron
radiation places \cite{/26/,/27/}. In case of an EMU the MU
paramter K is replaced by the Lorentz invariant normalized laser
strength $a_L$
\begin{eqnarray}
K = \frac{eB_U \lambda_U}{2 \pi mc } \ \to a_L = \frac{eE_L
\lambda_L}{2 \pi mc^2 } \label{eq(19)}
\end{eqnarray}
and the MU period length $\lambda_U$ corresponds to the EMU wavelength $\lambda_L$  via \cite{/26/}
\begin{eqnarray}
\lambda_U = \lambda_L / ( \cos\alpha_0 + 1 / \beta) \label{eq(20)}
\end{eqnarray}
i.e. for $\alpha_0 = 0$ (head-on collision) the energy $\omega_f $
of the MU fundamental in case of small $K$ is $\omega_f = 4
\gamma^2 \omega_L = 2 \gamma^2 \omega_U$ with $\omega_U = 2 \pi c
/ \lambda_U$. For an intensity $I_L = 10^{18}$~W/cm$^2 $  one
obtains for linearly polarized laser light an electric field
strength $E_L = (2 I_L / \epsilon_0 c )^{1/2} = 2.7 \ \cdot
10^{12}\ $V/m  and thus a laser parameter $a_L = 0.85$. We note
that this laser parameter induces an electron acceleration a which
corresponds to an Unruh-Hawking temperature \cite{/7/,/23/}
\begin{eqnarray}
T_{UH} = \hbar a / (2 \pi k_B c) = \hbar \omega_L a_L / (2 \pi k_B
) = 1900 \ \rm{K} \label{eq(21)}
\end{eqnarray}
and which is close to that for electrons in modern high energy
lepton storage rings \cite{/27b/} ($k_B=$Boltzmann constant).
Within the context of our discussion it might be indicated to make a short remark about thermal bathing of electrons in a storage ring. Due to radiative spin-flip transitions within the magnetic lattice of the storage ring electrons become polarized but the polarization stays always below 100\%. Bell and Leinaas \cite{/27b/} have attributed this depolarization to the thermal Unruh effect, thus replacing the standard interpretation in terms of the well-known Sokolov-Ternov effect in QED \cite{Soko}. But in a recent paper E.Akhmedov and D.Singleton \cite{Akhm}  have shown,``that the laboratary observer interprets the effect as the Solokov-Ternov effect while the non-inertial co-moving observer interprets it as the circular Unruh effect. Physically,these two effects are the same.''  Thus,also in this case,the Unruh effect has been well known for a long time under a different name and has even been experimentally observed \cite{Akhm}.\\

This is in accordance with statements from a very recent review article about the Unruh effect and its applicatons \cite{Cris}: 
 ``We shall emphasize that, although it is certainly fine to interpret laboratory observables from the point of view of uniformly accelerated observers, the Unruh effect itself does not need experimental confirmation any more than free quantum field theory does. It might happen, that some observables can be more easily computed and interpreted from the point of view of uniformly accelerated observers using the Unruh effect. This is a matter of convenience and not of principle''. In addition, several examples are given by the authors, where first each phenomenon is discussed by plain quantum field theory adapted to inertial observers and then it is shown, how the same observables can be reproduced from the point of view of Rindler observers with the help of the Unruh effect. Specifically they write that ``Sch\"utzhold et al.\cite{/7/} note that a Rindler photon seen to be scattered off a static charge by the Rindler observers should correspond to a pair of correlated Minkowski photons emitted from a uniformly accelerated charge as seen by the inertial observers. Although the correlated radiation of Minkowski photons could be explained by inertial observers using textbook quantum field theory, it is certainly interesting to understand these pocesses invoking the Unruh effect''.  The theory of double Compton scattering is nothing else than a description of the generation of Minkowski photon pairs in terms of textbook QED \cite{/13/}. Last but not least we quote a conclusion drawn by H.C.~Rosu \cite{Rosu} in his review article `Hawking-like Effects and Unruh-like Effects: Toward Experiments?': ``My feeling at the end of the survey is that actually the goal is not so much to try to measure a `Hawking' or a `Unruh' effect. Being ideal concepts/paradigms, what we have to do in order to put them to real work is to make them interfere with the many more `pedestrian' viewpoints.''\\

 The
value of $a_L$ compares also well with usual MU parameters K.
Since for $a_L > 1$ the transverse wiggling motion of the electron
within the EMU starts to deviate from an harmonic oscillation with
frequency $\omega_L$ due to the increasing importance of a
longitudinal Lorentz force which induces an additional oscillation
with frequency  $2 \omega_L$ (figure-of-eight motion \cite{/27/}),
harmonics in the radiation pattern are observed (multipole
radiation). But the analogon between MU and EMU radiation can even
be extended. For large MU parameters K the energy of the
fundamental is given by
\begin{equation}
\omega_f = 2 \gamma^2 \omega_U / \left(1+K^2+(\gamma
\theta)^2\right)\label{eq(22)}
\end{equation}
where $\theta$ is the emission angle. On the other hand all
expressions for single and double Compton scattering discussed
above have be obtained for electron plane waves in vacuum. But it
is well known that in strong electromagnetic fields the electron
is better described by Volkov states, since the 4-momentum of a
charged particle inside an electromagnetic wave is altered due to
continuous absorption and emission of photons. For a charged
particle with 4-momentum $p_{\mu}$ outside the field the effective
4-momentum $q_{\mu}$ (quasimomentum) inside the field is
\cite{/27c/}
\begin{equation}
q_{\mu}=p_{\mu} - \frac{a_L^2}{2(k \cdot p)}k_{\mu} \label{eq(23)}
\end{equation}
and thus $q_{\mu} q^{\mu}=-(1+a_L^2)\equiv-m_{eff}^2$. Due to the
interaction with the laser field the electron is ``dressed'', i.e.
it acquires an effective mass $m_{eff}>1$. Inspecting the formulas
of chapter II when all p's are replaced by q's, the most dramatic
effect occurs for the kinematics of eq.~(\ref{eq(1)}). Instead of
(\ref{eq(3)}) one obtains in case of inverse IDCS ($\gamma \gg 1$)
\begin{widetext}

\begin{eqnarray}
\omega_2 =   \frac{2 \gamma^2 \omega_L \left(1-\beta cos \alpha
\right) - \omega_1 \left( 1+a_L^2+(\gamma \theta_1')^2\right)
-2\gamma \omega_1 \omega_L \left( 1-cos \alpha \right)
}{1+a_L^2+(\gamma \theta_2')^2 + 2\gamma \omega_L \left(1- cos
\alpha \right)} \label{eq(24)}
\end{eqnarray}

\end{widetext}
and equivalently to eq.~(\ref{eq(13)})
\begin{equation}
\omega_{1max} \cong \frac{2 \gamma^2 \omega_L \left( 1-\beta cos
\alpha \right)}{1+a_L^2 + (\gamma \theta_1')^2} \label{eq(25)}
\end{equation}
which in case of head-on collisions ($\alpha = \pi$) becomes
\begin{equation}
\omega_{1max} = \frac{4 \gamma^2 \omega_L}{1+a_L^2 + (\gamma
\theta_1')^2} \label{eq(26)}
\end{equation}
Eq.~(\ref{eq(26)}) establishes further more the equivalence of
EMU-and MU-radiation, see eq.~(\ref{eq(22)}). This nonlinear QED
mass-shift effect \cite{/27/,/29/} is not small, since for
$a_L=0.85$ it would predict a $40\%$ reduction of $\omega_{1max}$.
Strictly spoken, this holds for a flat-top shaped laser pulse with
a unique strength $a_L$ only. In case of a Gaussian beam one has
an intensity and hence an $a_L$-distribution. During its encounter
with the laser pulse the electron will emit radiation with a
variety of mass-shifts, resulting in a so called ``ponderomotive
broadening'' \cite{/34a/} which would strongly effect the single
Compton peak in Fig.~7 by increasing its width by approximately
$40\%$. A corresponding shift in case of the 12.4 keV radiation,
which we have assumed for fixed targets, turns out to be
negligible small. For a lateral extension of the photon beam
$\sigma_{t\gamma}=35 \mathrm{\mu m}$, $10^{12}$ photons/pulse and
a 100 fs pulse duration \cite{/18/}, we obtain a pulse intensity
of $\mathrm{I}=5\cdot 10^{14} \mathrm{W/cm^2}$ and thus a
normalized XFEL-strength $\mathrm{a_L}=2\cdot 10^{-6} \ll 1$. If
it would be possible to focus the XFEL beam down to the
diffraction limit $\sigma_{t\gamma} = \lambda_L$-up to now with a
method not known - one could even reach $\mathrm{a_L}=0.7$. While
there are unambiguous indications of the generation of harmonics
in inverse single Compton scattering \cite{/34b/} a corresponding
signature for the associated mass-shift discussed above has, to
our knowledge, not yet been verified. Therefore, we have refused
to include this effect in our analysis.\\
From such an analogon two conclusions can be drawn: \\
1) it might be useful to look for higher harmonics in inverse IDCS
similar to those of inverse single Compton scattering (i.e.
nonlinear Thomson scattering \cite{/27/,/29/}). This, in fact,
would be a multi-photon process, i.e. the four vector k in
(\ref{eq(1)}) should be replaced by nk where n is the number of
simultaneously absorbed laser photons \cite{/27c/}.\\
2) one may look for entangled photons in energy regions between
the peaks of MU-FEL radiation.  Due to the excellent brilliance of
such a source one might even speculate about some transverse
(spatial) coherence of the photon pairs, i.e.~a correlation of the
amplitudes of pair-states in transverse space of the FEL-beam
\cite{/30/}, thus realising a kind of ``photon pair
laser''\cite{/7b/}. In nonlinear optics such a multiple photon
pair-state is called a one- or two-mode squeezed vacuum state
\cite{/33/,Ahn}.

\section{Conclusion}
It has been demonstrated that IDCS is well suited as a source for
entangled X-ray photons, though we admit that the increase of the
yield compared to that of CDCS is not dramatic.   We remark that
the cross section function  X of (\ref{eq(5)}) has been obtained
by averaging over the initial and summing over the final
polarization  vectors, i.e.~the cross section holds for
unpolarized radiation only. It would be therefore  desirable to
repeat the calculation for polarized incident light. In absence of
such a calculation we argue as follows: in the rest frame of the
electron the relative inelasticity for 180$^o$ backscattering is
$\Delta \omega^* / \omega^* = 2 \omega^* $. Thus, if $2 \omega^*
<< 1 $ we expect that scattering is well described by Thomson
scattering. Since $\omega^* = 2 \gamma \omega_L$  the condition
reads $\omega_L << 1 / (4 \gamma)$ . For  $\gamma = 300$ we have
$\omega_L << 10^{-3}$, which is fulfilled in our scenario.  But it
is well known that in case of Thomson scattering a 100\% linearly
polarized incoming photon beam remains 100\% polarized,
independently of the scattering angle. Since the Stokes parameter
for linear polarization is a Lorentz invariant this holds for the
lab frame also. Therefore, for inverse single Compton scattering
the final photons will be 100\% linearly polarized. We suspect a
similar behavior for IDCS, i.e. the polarization vectors of both
final photons point in the same direction as that of the initial
photon yielding maximal entanglement.\\
Finally, we comment on some notions. Since the total cross section
of ordinary Thomson scattering is about $r_0 ^2 $, this result can
also be obtained by classical physics. Thus, Larmor radiation
which, of course, includes undulator radiation and its harmonics
might be called classical radiation.  In contrast, IDCS (whose
cross section is proportional to $\alpha_{QED} r_0 ^2 $), or Unruh
radiation \cite{/7/,/7b/} can be termed quantum radiation. Certainly, the Unruh effect itself is an important test ground for theoreticians (L.C.B.~Crispino et al.\cite{Cris} call the effect a {\it theoretical laboratory}) to investigate quantum field theories in non-inertial reference frames, but it seems questionable if it should play an equivalent role for experimentalists. \\
I am indebted to Bernhard Adams and Ralf Sch\"utzhold for valuable
discussions.

\end{document}